\documentclass[pre,twocolumn,aps,superscriptaddress, nofootinbib]{revtex4-2}
\usepackage{amsmath}
\usepackage{amssymb}
\usepackage{graphicx}
\usepackage{subcaption}
\usepackage[usenames,dvipsnames]{color}
\usepackage{braket}
\usepackage{bm}
\usepackage{times}
\usepackage{hyperref}
\usepackage{pdfpages}
\usepackage{float}
\usepackage{lipsum}
\usepackage[dvipsnames]{xcolor}
\usepackage{comment}
\usepackage{ragged2e}
\usepackage[normalem]{ulem}

\usepackage{xr}

\hypersetup{
  colorlinks=true,
  citecolor=blue,
  linkcolor=blue,
  urlcolor=blue}
\usepackage{tikz}
\usetikzlibrary{decorations.markings}
\usetikzlibrary{decorations.pathmorphing,snakes}
\usetikzlibrary{arrows.meta}
\tikzset{middlearrow/.style={
        decoration={markings,
            mark= at position 0.5 with {\arrow{#1}} ,
        },
        postaction={decorate}
    }
}

\def\be{\begin{equation}}
    \def\ee{\end{equation}}
\def\bea{\begin{eqnarray}}
    \def\eea{\end{eqnarray}}
\def\bfg{\begin{figure}[H]}
    \def\efg{\end{figure}}

\makeatletter
 \AtBeginDocument{\let\LS@rot\@undefined}
\makeatother

\begin{document}

\title{{Emergence of Local Ordering and Mesoscale Giant Number Fluctuations in Active Turbulence}}

\author{Kirti Kashyap}
\email{kirtikashyap26@gmail.com}
\affiliation{Department of Physics, Indian Institute of Technology Hyderabad, Hyderabad, India}

\author{Kolluru Venkata Kiran}
\email{kiran8147@gmail.com}
\affiliation{Universit\`e C\^ote d'Azur, CNRS, Institut de Physique de Nice (INPHYNI), 17 rue Julien Laupr\^etre, 06200 Nice, France}

\author{Anupam Gupta}
\email{agupta@phy.iith.ac.in}
\affiliation{Department of Physics, Indian Institute of Technology Hyderabad, Hyderabad, India}

\begin{abstract}

We study spatiotemporal chaos in two-dimensional dense active suspensions using a generalized hydrodynamic model. Increasing activity induces a structural transition marked by the formation of intense vortices and giant number fluctuations at the mesoscale. The flow self-organizes into locally polar-ordered regions coexisting with chaotic domains, producing a bimodal velocity distribution and enhanced correlations. This mixed-state morphology underlies the universal statistical behavior observed beyond a critical activity threshold. Reducing the instability timescale yields similar transitions, showing that both activity and instability act as control parameters for pattern formation. An energy-based order parameter derived from the system’s budget quantifies and unifies these structural transitions across the phase space of activity and instability timescales.

\end{abstract}

\maketitle 
\textit{Introduction}---The collective behavior of active particles gives rise to complex and intriguing phenomena, including large-scale coherent motion in flocks, mesoscale turbulence in bacterial suspension, and defect-driven chaos in active liquid crystals \cite{alert2022active,marchetti2013hydrodynamics,toner1998flocks,toner2005hydrodynamics,giomi2013defect,james2021emergence,ngo2014large,guillamat2016control,wensink2012meso,sokolov2007concentration,martinez2020combined,guillamat2017taming,wei2024scaling,rorai2022coexistence}.
The spatiotemporal structures observed in dense suspensions of active matter qualitatively resemble those found in passive turbulence systems\cite{wensink2012meso,alert2022active,martinez2020combined,wolgemuth2008collective,dombrowski2004self,sokolov2007concentration,doostmohammadi2019coherent}. Extensive studies have been conducted to investigate whether active turbulence shares similarities with or deviates from inertial turbulence. These investigations often focus on identifying universal features, such as power-law scaling and intermittency, which are hallmarks of turbulent systems~\cite{cp2020friction,martinez2021scaling,mukherjee2023intermittency,kiran2023irreversibility,Pandit_2025}. Interestingly, the statistical properties at high activity resemble those of inertial turbulence—exhibiting scaling and intermittency:
recent studies demonstrated the emergence of universal scaling and intermittency in the energy spectrum beyond a critical activity threshold~\cite{mukherjee2023intermittency,kiran2025onset}.

While these statistical features closely resemble those of inertial turbulence, their physical origin in active systems remains unclear. In particular, it is not known whether the universal scaling observed at high activity arises from fully chaotic dynamics or from a more complex coexistence of ordered and disordered regions.

In this work, we demonstrate that beyond a critical activity threshold, the flow field self-organizes into locally ordered domains coexisting with chaotic regions, and that this mixed-state structure underlies the universal behavior observed in active turbulence. By systematically varying the activity and instability timescale, we show that both serve as control parameters governing the onset of local ordering and the associated structural transitions.
However, there is also evidence pointing to structural changes in the vorticity field itself with increasing activity~\cite{ramaswamy2016activity,linkmann2019phase,linkmann2020condensate,mukherjee2021anomalous,puggioni2022giant,gautam2024harnessing,ariel2015swarming}. This highlights the need for a deeper understanding of how these emergent structures contribute to the observed universal and nonuniversal features of active turbulence. Gaining insight into these structural changes and flow dynamics is important for explaining various processes like nutrient mixing and molecular transport in biological and ecological systems \cite{sokolov2009reduction,kurtuldu2011enhancement,dombrowski2004self,hernandez2005transport}.

\begin{figure*}
    \centering
    \includegraphics[width=1.0\linewidth]{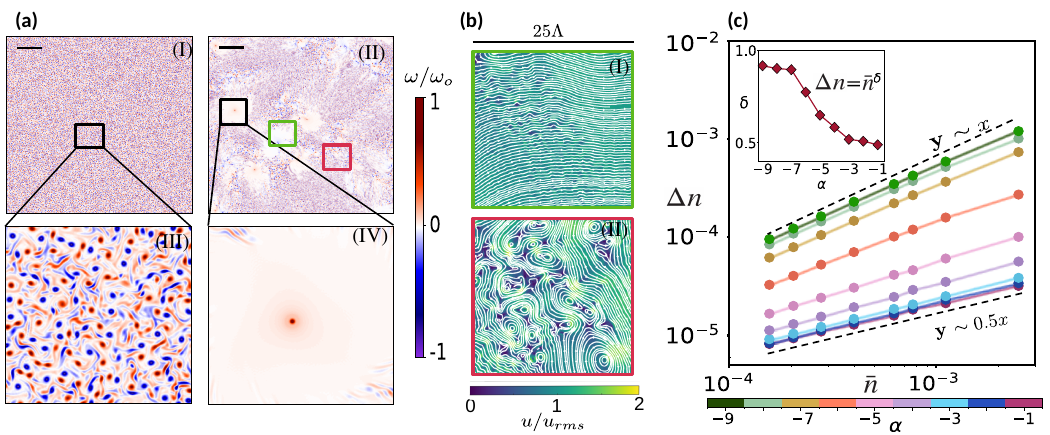}
    \caption{\justifying \small{The vorticity field ($\omega$) is shown for two different activity levels: (a I) $\alpha = -1$, and (a II) $\alpha = -9$ at $ t =85\tau_{\Gamma}$ with $\tau_{\Gamma}=0.4$. The scale bar in both panels corresponds to $25\Lambda$. Panels (a III) and (a IV) present zoomed-in sections of the respective vorticity fields, with magnified regions chosen to be of the same order as the scale bar. The vorticity snapshot at higher activity displays spatial inhomogeneities, characterized by intense vortex regions coexisting with areas of negligible vorticity. The color bar is normalized by the maximum vorticity $\omega_0$ of the respective $\alpha$}. (b I) and (b II) display the velocity streamlines superimposed on the velocity magnitude within a subdomain of size $25\Lambda$, corresponding to the regions marked by the red and green boxes in (a II). (c) Standard deviation $\Delta n$ of the number of vortex centers within subregions is plotted against the corresponding mean $\bar{ n}$ as the subregion size $\ell$ increases. The scaling behavior $\Delta n \sim \bar{n}^\delta$ reveals the presence of giant number fluctuations for $\alpha < \alpha_c$. Inset: the scaling exponent $\delta$ is plotted as a function of the activity parameter $\alpha$, showing that fluctuations become increasingly pronounced (larger $\delta$) with higher activity.} 

    \label{fig:vorcity_field}
    \end{figure*}

We employ hydrodynamic models developed to describe the complex, chaotic flows in dense, quasi-two-dimensional bacterial suspensions~\cite{wensink2012meso,mukherjee2023intermittency,kiran2025onset}. These models have also been used to study turbulence-like behavior in dense suspensions of \textit{Bacillus subtilis} \cite{alert2022active,Pandit_2025}. While previous studies have primarily focused on statistical signatures such as energy spectra or intermittency, the energetic underpinnings of these transitions remain poorly understood. We introduce an energy-based order parameter that provides a unifying framework to distinguish ordered and disordered regimes in active turbulence, linking structural organization directly to the system’s underlying energy balance and examining how the flow field reorganizes as activity and instability compete. Increasing activity leads to a strikingly uneven distribution of intense vorticity centers in space and time (Fig~\ref{fig:vorcity_field}(a)). Tracking the number of vortex centers reveals emergence of giant number fluctuations at the mesoscale, i.e., at length scales larger than the characteristic vortex size but smaller than the system size, hinting at an underlying structural transition in the flow. To probe the underlying origin of this transition, we examine the spatial velocity correlations and observe a marked increase in correlation length, along with the onset of local polar ordering. Interestingly, these locally ordered regions do not take over the whole system; instead, they coexist with disordered, chaotic zones. 
Furthermore, we find that tuning the timescale of the linear instability, introduced to model intrinsic instabilities in dense active suspensions, can alter the system’s phase even at fixed activity levels.
This effectively shifts the critical activity threshold for structural transitions.

\textit{Generalized hydrodynamic model}---
Following the hydrodynamic frameworks proposed for dense bacterial suspensions~\cite{wensink2012meso}, we describe the system using an incompressible velocity field $\mathbf{u}(\mathbf{x}, t)$ that evolves according to : 
\begin{equation}
\resizebox{0.9\columnwidth}{!}{$ 
\begin{aligned}
\frac{\partial \mathbf{u}}{\partial t}+\lambda_0 \mathbf{u} \cdot \nabla \mathbf{u} &= -\nabla p-\left(\alpha+\beta|u|^2\right) \mathbf{u}+\Gamma_0 \nabla^2 \mathbf{u}-\Gamma_2 \nabla^4 \mathbf{u} \\
\nabla \cdot \mathbf{u} &= 0
\end{aligned}
$}
\label{eq:TTSH}
\end{equation}

$p(\boldsymbol{x},t)$ is the pressure field. The coefficient $\lambda_0$  encodes the effects of different swimmer types: $\lambda_0>1$ corresponds to pusher-type swimmers, while $\lambda_0<1$ corresponds to puller-type swimmers. The Toner-Tu coefficients $(\alpha,\beta)$ describe the polar ordering in self-propelled particle systems~\cite{toner1998flocks,toner2005hydrodynamics}: for values of $\alpha < 0 (>0)$ the term $\alpha \boldmath{u}$ injects (dissipates) energy and $\beta>0$ suppresses energy buildup at large length scales. 
$\alpha <0$ and $\beta$ define a characteristic velocity scale, $v = \sqrt{|\alpha| / \beta}$, for polar-ordered motion. The ($\Gamma_0,\Gamma_2$) term has the structure of the Swift-Hohenberg equation \cite{swift1977hydrodynamic}. With $\Gamma_0 < 0$ and $\Gamma_2 > 0$, this term inherently drives the system toward an instability at the length scale 
$\Lambda = 2\pi \sqrt{\tfrac{2\Gamma_2}{|\Gamma_0|}}$, velocity $v_0 = \sqrt{\tfrac{|\Gamma_0|^3}{\Gamma_2}}$ and    
the corresponding instability growth timescale is 
$\tau_{\Gamma} = \tfrac{\Lambda}{v_0}$.
It is the interplay between these terms that leads to chaotic, self-sustaining, turbulent-like flow patterns.

\textit{Simulation details}--- We solve Eq. (1) using a de-aliased pseudo-spectral method with periodic boundary conditions. Simulations are performed in a two-dimensional square domain of size, L$=160$ with $4096^2$ grid points, ensuring adequate resolution of fine-scale features. Time integration employs the second-order integrating factor scheme (IFRK2) with a time step $\Delta t = 2\times10^{-4}$. The simulations were initialized with a random vorticity field, and in all cases, we set the parameters fixed (except $\alpha$) unless otherwise stated, $\lambda_0= 3.5$, $\beta = 0.5$, $\Gamma_2 = 9\times10^{-5}$, $\Gamma_0  = -4.5\times10^{-2}$. These values are chosen based on previous studies that align with experimentally accessible regimes~\cite{kiran2023irreversibility}. For more details, see section~\ref{numerical_methodology}. To investigate the role of activity, we systematically vary the parameter $\alpha$ between -9 to -1. 
Using average velocities observed in experiments on B. \textit{subtilis} at normal oxygen concentration, $\sim 25 \mu m/s$ ~\cite{wensink2012meso,cp2020friction,kiran2023irreversibility}, our simulation parameters map to physical $v_{rms}$ in the range $4.5 - 70 \mu m/s$, consistent with experimental observations~\cite{liu2012multifractal}, with the full correspondence detailed in ~\ref{numerical_methodology}.

\begin{figure*}[htbp]
    \centering
    \includegraphics[width=1\linewidth]{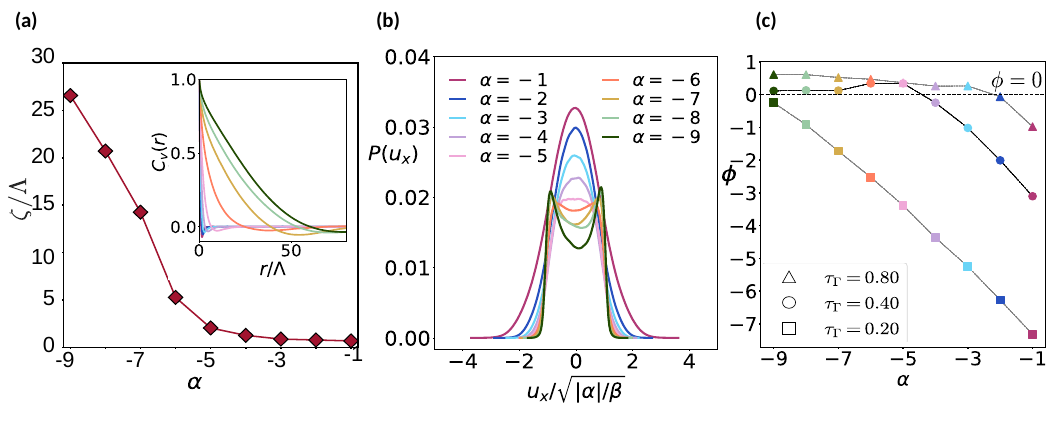}  
    \caption{\justifying \small{(a) The normalized velocity correlation length $\zeta/\Lambda$ versus $\alpha$. \textit{Inset:} The spatial velocity correlation function, defined as  $C_v(r) = \langle \mathbf{u}(\mathbf{r}_0) \cdot \mathbf{u}(\mathbf{r}_0 + \mathbf{r}) \rangle$ is shown as a function of distance $r$ for increasing values of $|\alpha|$. $\zeta$ is extracted by fitting the decay of $C_v(r) \sim e^{-r/\zeta}$. For $\alpha > \alpha_c$, $\zeta/\Lambda \sim 1$, whereas it increases significantly for $\alpha < \alpha_c$, indicating the emergence of long-range velocity correlations in the high-activity regime. (b) Probability distribution function of the normalized velocity component $u_x/\sqrt{|\alpha|/\beta}$ for different activity values $\alpha$. A transition from a unimodal to a bimodal distribution is observed as $\alpha$ decreases beyond the critical point $\alpha_c$. (c) The order parameter $\phi = E_{\rm active} - E_{\Lambda} - E_{\rm kin}$ is plotted as a function of $\alpha$ for three values of $\tau_\Gamma$ obtained by varying $\Gamma_2$. The baseline case with $\tau_{\Gamma} = 0.4$ ($\Gamma_2 = 9.0\times10^{-5}$, as specified in the parameters) is shown by circles, while increased and decreased $\tau_{\Gamma}$ are represented by triangles and squares, respectively. For intermediate and larger $\Gamma_2$, $\phi$ remains positive below the corresponding critical activity $\alpha_c$ and becomes negative above it, marking the transition. In contrast, for the smallest $\Gamma_2$, $\phi$ stays negative across all $\alpha$, and no transition point $\alpha_c$ is observed.}}
    \label{fig:vel_corr}
\end{figure*}

\textit{Spatial distribution of vortex centers}---
We show in Fig.~\ref{fig:vorcity_field} (a) the plots of vorticity field $\omega(\mathbf{x}, t) = \nabla \times \mathbf{u}(\mathbf{x}, t)$ for $\alpha = -1$ (I) and $-9$ (II). The field $\omega$ undergoes a structural transformation as $\alpha$ decreases, with large-scale spatial inhomogeneities becoming more prominent at lower $\alpha$ or, equivalently, higher activity. A similar feature has also been observed in earlier studies of active turbulence with increasing activity \cite{mukherjee2021anomalous,puggioni2023flocking}. 
These inhomogeneities are evident from the large-scale vortical structures as seen in Figs.~\ref{fig:vorcity_field}(a III) and ~\ref{fig:vorcity_field}(a IV). In active matter systems, such large-scale spatial variations are a hallmark of giant number fluctuations, wherein the standard deviation ($\Delta n$) of the number of observables --- in this case, vortex centers ($n$) --- grows anomalously large and scale with the average as $\Delta n \sim \bar{n}^\delta$, with the exponent $\delta$ reaching values close to unity in two dimensions~\cite{chate2020dry,narayan2007long,fily2012athermal}. In the asymptotic limit of $\rm n$, $\delta=0.5$ for systems in equilibrium and away from continuous phase transitions. In analogy, we find giant number fluctuations of the vortex centers in active turbulence. To quantify these fluctuations, we measure the average number of vortices, $\bar{n}$, and the corresponding standard deviation, $\Delta n$, within a subregion of size $l^2$. 
As $\alpha$ decreases, $P(n/\bar{n})$ broadens, revealing stronger fluctuations in the number of vortex centers (Fig.~\ref{fig:vor_distribution}).
 In Fig.~\ref{fig:vorcity_field}(c) we plot $\Delta n$ versus $\bar{ n}$ for different activities and find that $\Delta \rm n \sim \bar{ n}^{\delta}$, with 
 $\delta \simeq 0.5$ for $\alpha \geq-5$ and $\delta > 0.5$ for  $\alpha <-5$. The values of $\delta$ approach 1 for large activity values, see inset of Fig.~\ref{fig:vorcity_field}(c). We find that giant number fluctuations persist up to a subregion of size order $20 \Lambda$ $\sim \mathcal O$(correlation length) (Fig.~\ref{fig:vel_corr}(a)), beyond which number fluctuations start decaying toward homogeneity. Notably, the transition of $\delta$ at a critical value of the activity parameter, $\alpha_c = -5$, matches with the previously identified critical activity by the group of Ray \cite{mukherjee2021anomalous,mukherjee2023intermittency} through the onset of universality in energy spectra, increased intermittency, and other hallmarks of turbulence. Note that with increasing activity, although the vorticity at individual vortex centers increases, the total number of such centers decreases~\ref{fig:num_vor}(b). \\
Giant number fluctuations are typically attributed to long-range correlations in density fluctuations~\cite{dey2012spatial,toner2024physics,toner2005hydrodynamics,fily2012athermal}, seen as a divergence of the structure factor in the long-wavelength limit. However here, such density fluctuations are absent because the fluid is incompressible. Nevertheless, one can define an effective number density for vortex centers, $\rho_l(\bm{x},t)=n/l^2$. The static structure factor of $\rho_l$ grows with decreasing $k$ and peaks at a finite wavenumber (corresponding length is $\sim50\Lambda$, for $\alpha=-6$), indicating the presence of giant number fluctuations at a mesoscale (see section~\ref{structure factor}), Fig~\ref{fig:static_structure_factor}. Consistently, the structure factor remains finite in the long-wavelength limit, reflecting the fact that these giant number fluctuations arise at a mesoscale and not at system sizes $~\sim400\Lambda$. Furthermore, when giant number fluctuations are absent, corresponding to $\alpha>-5$, the structure factor does not have a characteristic peak and saturates to a constant value as $k\to0$ (see Fig~\ref{fig:static_structure_factor}), indicating homogenization of fluctuations in $\rho_l$ \cite{padhan2025suppression}.\\
 The methodology for identifying vortex centers is provided in section \ref{extract_vor_cenerts}. While the fluctuations in number reflect statistical deviations, a more structural perspective on the spatial inhomogeneity of vortex centers is offered by Voronoi tessellation. The results of this analysis are consistent with the emergence of giant number fluctuations at high activity levels and show a transition at $\alpha_c = -5$ (see section~\ref{voronoi_dist}). 

\textit{Emergence of local ordering}---
The emergence of giant number fluctuations raises the question of what structural reorganization in the velocity field underlies this transition.
To address this, we closely examine the structure of the vorticity and velocity fields. For $\alpha<\alpha_c$, there are a few large vortical regions of the size $~\sim 25\Lambda$ characterized by low or nearly vanishing vorticity (see Fig.~\ref{fig:vorcity_field}(a IV)) interspersed with vortical regions of the size $\mathcal{O}(\Lambda)$. In Fig~\ref{fig:vorcity_field}(b I), we plot the velocity streamlines corresponding to these large vortical regions of vanishing vorticity and note that they are highly aligned, indicating local ordering. This contrasts sharply with other regions of the same field Fig~\ref{fig:vorcity_field}(b II), where the flow remains disordered and lacks directional alignment. Furthermore, in Fig~\ref{fig:vel_corr}(a) we plot velocity correlation function $C_v(r) = \langle \mathbf{u}(\mathbf{r}_0) \cdot \mathbf{u}(\mathbf{r}_0 + \mathbf{r}) \rangle$ versus $r$ and see that there is a sharp transition followed by a monotonic increase in correlation length, $\zeta$, with decreasing $\alpha$, for $\alpha<\alpha_c$. For sufficiently negative $\alpha$ (e.g., $\alpha \lesssim -6$), the spatial velocity correlation length reaches up to $\mathcal{O}(10\Lambda)$ and for $\alpha=-9$ we find $\zeta\simeq 25\Lambda$, consistent with the lengths of the large vortical regions seen in Fig.~\ref{fig:vorcity_field} (a II). For $\alpha>\alpha_c$, we note from Fig.~\ref{fig:vorcity_field} (a I) and (a III), the largest vortical structures are $\sim\mathcal{O}(\Lambda)$ and velocity streamlines are chaotic throughout domain, see Fig.~\ref{fig:num_vor}(a). Musacchio and group~\cite{puggioni2022giant} have reported a similar trend, however, they observe a saturation of the correlation length to an asymptotic value for high activities, rather than the monotonic growth found in our system. This saturation is attributed to the confined geometry employed in their study. 

Signatures of local ordering can also be seen in the probability distribution function (PDF) of the velocity field: we plot in Fig. \ref{fig:vel_corr}(b) the  PDF of a component of the velocity field for different activities and observe that the PDF transitions from a unimodal distribution centered at $\langle u \rangle = 0$  for $\alpha > \alpha_c$ to a bimodal distribution for $\alpha<\alpha_c$. The two peaks in the bimodal state progressively approach $\pm \sqrt{\frac{\left| \alpha \right|}{\beta}}$, consistent with the predictions of the Toner-Tu model for self-propelled active systems~\cite{toner1998flocks}. This highlights competition between active alignment-induced ordering Toner-Tu terms and spatial modulation introduced by higher-order gradient terms.

To gain a deeper understanding of the local ordering that emerges along with the large vortical structure in Fig.~\ref{fig:vorcity_field}(a IV), we analyze the velocity field around the center of this vortex. We represent the velocity field by shifting the origin to the vortex center and expressing it in polar coordinates as follows:
\[
\mathbf{u}(r) = u_r(r)\,\hat{\mathbf{r}} + u_\theta(r)\,\hat{\boldsymbol\theta}
\]
where $u_r$ and $u_\theta$ denote the radial and azimuthal velocity components, respectively. Our numerical simulations show that $u_{r}\ll u_{\theta}$ and
$u_\theta \sim \sqrt{|\alpha|/\beta}$ for $r > \Lambda$ (Fig.~\ref{fig:vel_comp_alpha_7}), consistent with the asymptotic solution~\cite{puggioni2023flocking}. To investigate the behavior near the vortex center, i.e., in the regime $r \ll \Lambda$, we assume a power-law ansatz $u_\theta \sim r^a$.
In this regime, the dynamics are dominated by the linear scaling $u_\theta \sim r$ (see detailed discussion in Sec.~\ref{Polar_form}). However, due to resolution limits, the small-$r$ behavior cannot be reliably extracted from the simulations and thus remains beyond the scope of the present work. Although this scaling may be related to the dynamics of vortex defects reported in the incompressible Toner–Tu (ITT) system~\cite{rana2020coarsening}, establishing this connection requires further study, which we leave for future investigation.

The global phase behavior is not dictated by a single control parameter but instead emerges from a balance between multiple energetically distinct contributions. The first is the energy associated with activity-driven motion, quantified as
$
E_{\rm active} = \frac{v^2}{2}.
$
The second arises from the destabilizing active stresses introduced by the generalized Swift–Hohenberg-like term, which generates intrinsic instabilities and injects energy into the system at intermediate length scales. The energy associated with this contribution is given by
$
E_{\Lambda} = \frac{|\Gamma_0|^3}{\Gamma_2}.
$
The third component is the kinetic energy of the flow,
$E_{\rm kin} (\equiv \int_{\mathbf{x}\in L^{2}} \frac{1}{2}\mathbf{u}\cdot\mathbf{u} d\mathbf{x}$).
This energetic framework allows us to define an effective order parameter, $\phi$, that captures the interplay between these different forms of energy input and transfer.
\[
\phi = E_{\rm active} - E_{\Lambda} - E_{\rm kin},
\]
We hypothesize that the relative magnitudes of $E_{\rm active}$, $E_{\Lambda}$, and $E_{\rm kin}$ determine the qualitative state of the system --- whether it exhibits local ordering identified by long correlation lengths or remains in a disordered phase.
For $\alpha < \alpha_c$, the order parameter $\phi$ becomes positive, indicating that the active energy $E_{active}$
dominates over other contributions. This leads to a breakdown of homogeneity and the emergence of local ordering. In contrast, $\phi<0$ corresponds to a homogeneous, disordered state where active energy is insufficient to induce such organization, Fig. \ref{fig:vel_corr}(c).

\textit{Instability growth timescale as an alternative control parameter}--- 
Having explored the role of the activity parameter $\alpha$, which sets a characteristic timescale $\tau_{\alpha} = 1/|\alpha|$ for the growth of polar order, we now turn to a complementary perspective: fixing $\alpha$ and varying the timescale associated with the growth of instabilities, $\tau_{\Gamma}$. In previous sections, $\tau_{\Gamma}$ was kept fixed while $\alpha$ was varied to investigate its impact on the emergence of large-scale structures. However, since both parameters control the growth rate of perturbations~\cite{wensink2012meso}, we now ask whether varying $\tau_{\Gamma}$ at fixed $\alpha$ can produce effects analogous to varying $\alpha$ at fixed $\tau_{\Gamma}$.

\begin{figure}
    \centering
    \includegraphics[width=0.7\linewidth]{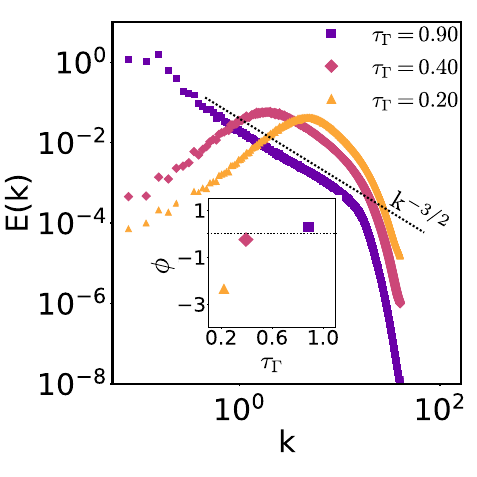}
    \caption{\justifying \small{The energy spectra by varying the instability growth timescale $\tau_{\Gamma}$ by varying $\Gamma_0$ for fixed $\alpha = -4$, $\Gamma_2=9\times10^{-5}$, $\beta=0.5$ and $\lambda_0=3.5$. Inset: The corresponding energy-based order parameter $\phi$ is plotted as a function of $\tau_{\Gamma}$, showing a sign reversal that is consistent with a transition. }} 
   \label{fig:spectra_gamma}
\end{figure}

We fix $\alpha = -4$, below the critical threshold $\alpha_c$, and vary the instability growth timescale, $\tau_{\Gamma} $ by adjusting $|\Gamma_0|$. For increased $\tau_{\Gamma}$ (i.e., smaller $|\Gamma_0|$), the energy spectra shown in Fig.~\ref{fig:spectra_gamma} exhibit a power-law scaling of the form $\rm E(k) \sim k^{-3/2}$, qualitatively similar to that observed for large $|\alpha|$ ~\cite{mukherjee2023intermittency} (Fig~\ref{fig:vrms_spectra}(b)) . Conversely, reducing $\tau_{\Gamma}$ by increasing $|\Gamma_0|$ produces spectra resembling those at smaller $|\alpha|$, (increased $\tau_{\alpha}$). Notably, the order parameter $\phi$ sharply captures this transition, changing sign in correspondence with the shift in flow structure (Fig.\ref{fig:spectra_gamma}, inset). \\

To further assess the generality of the order parameter, we extended the analysis to two additional representative values of $\tau_\Gamma$ ($\Gamma_2$), as shown in Fig.~\ref{fig:vel_corr}(c). For decreased $\tau_\Gamma$ (corresponding to $\Gamma_2 = 3.6\times10^{-5}$), the system remains disordered with $\phi < 0$ for all values of $\alpha$ considered, and no local ordering emerges. In contrast, for an increased $\tau_\Gamma$ (larger $\Gamma_2 = 1.8\times10^{-4}$), the system undergoes a transition to a locally ordered state with $\phi > 0$ already at much lower activity levels ($\alpha = -2$). These results demonstrate that both $\alpha$ and $\tau_\Gamma$ act as key control parameters for the system’s dynamics, jointly determining the onset of local ordering. Correspondingly, the energy spectra exhibit distinct signatures, reflecting the redistribution of energy across scales as the system transitions between disordered and locally ordered states (Fig.~\ref{order_param_add}(c)). In parallel, the velocity statistics undergo a qualitative change: while the distribution remains unimodal in the disordered regime, increasing $\tau_\Gamma$ drives the system into a locally ordered state characterized by a bimodal velocity distribution (Fig.~\ref{order_param_add}(b)).

\textit{Discussion}--- 
We identify a previously unrecognized structural transition in active turbulence, marked by the simultaneous onset of local ordering and giant number fluctuations. 
The Toner-Tu terms promote local ordering due to long-range polar alignment \cite{toner1998flocks,toner2005hydrodynamics}. In contrast, the Swift-Hohenberg term introduces a counteracting influence, driving the system toward spatially chaotic structures \cite{swift1977hydrodynamic}. 
Increasing activity enhances local polar order and consequently increases the velocity correlation length. In the high-activity regime, signatures of local ordering are reflected in the single-point velocity statistics, which become bimodal. This regime also coincides with the onset of anomalous diffusion in bacterial turbulence, reported in simulations \cite{mukherjee2021anomalous} and observed experimentally \cite{ariel2015swarming,gautam2024harnessing}. This suggests a potential connection between transport behavior and the underlying velocity statistics.
This transition is sensitive to the strength of the higher-order derivative terms $\Gamma_0$ and $\Gamma_2$, hence $E_{\Lambda}$ and is robustly captured by the order parameter $\phi$, which switches sign in accordance with the onset of structural inhomogeneity. Our analysis reveals that the velocity field transitions into a state in which ordered vortical domains, characterized by large correlations, interact with disordered fluctuations, leading to the observed universal behavior at high activity.
Beyond serving as a quantitative marker, the energy-based order parameter encapsulates this competition among active driving, instability, and kinetic dissipation, providing a unified energetic criterion for pattern selection in active turbulence. We anticipate that this framework will help connect statistical universality to structural organization in other active and driven dissipative systems~\cite{meunierPRL2025}. 
The predicted transition could be probed in dense suspensions of fast-swimming bacteria or active colloids with tunable alignment by measuring the emergence of bimodal velocity statistics and correlated vortex clustering.
The recent 3D simulations of the Toner-Tu-Swift-Hohenberg model reveal stable flocking regimes at high activity that are absent in 2D~\cite{perlekar2026flocking}, suggesting that the coexistence of ordered and chaotic states identified here may acquire a qualitatively new character in three-dimensional active systems.
 Together, these findings provide a unified framework linking structural, statistical, and energetic signatures of turbulence in active matter.

\textit{Acknowledgements} --- We thank S. Mukherjee, S.S. Ray, P. Perlekar, S. Musacchio, and S. Islam for helpful discussions. These simulations were performed on the IITH NVIDIA V100 GPU cluster, Param Seva supercomputers through the National Supercomputing Mission, and the IITH Kanad clusters.
KK acknowledges the TCS Foundation for financial support through the Research fellowship program (TCS/17/23-24/P49).
KVK is grateful for funding from the Simons' Collaboration on Wave Turbulence (award No. 651471).
A. G. acknowledges SERB-DST (India) Projects MTR/2022/000232, CRG/2023/007056-G, DST (India) grant no. DST/NSM/R\&D\_HPC\_Applications/2021/05 and grant no. SR/FST/PSI-215/2016, and IITH for Seed Grant No. IITH/2020/09 for financial support.

\bibliography{reference}
 \bibliographystyle{unsrt}

%
%



\clearpage
\onecolumngrid
\definecolor{orange}{RGB}{255,127,0}
\definecolor{blue2}{RGB}{33,114,173}
\renewcommand{\theequation}{S\arabic{equation}}

\renewcommand{\thefigure}{S\arabic{figure}}  
\renewcommand{\figurename}{Supplementary Figure}  
\renewcommand{\thesection}{S\arabic{section}}  
\setcounter{figure}{0}  


\section*{Supplemental Material} 

\section{Numerical Methods}
\begin{figure} [h!]
    \centering
    \includegraphics[width=0.7\textwidth]{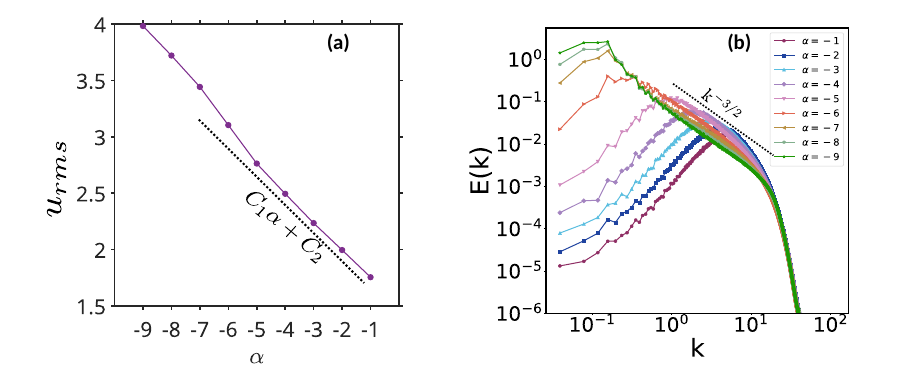}
    \caption{\small{(a) $u_{rms}$ increases with activity, (increasing $-\alpha$). (b)Energy spectra for different $\alpha$, showing power scaling for $\alpha < \alpha_c $.}}.
    \label{fig:vrms_spectra}
    \end{figure}

{\label{numerical_methodology}}
To numerically solve Equation~(\ref{eq:TTSH}) in the main text, we have used vorticity-stream function formalism \cite{sb2000turbulent}.
Taking curl of Equation~(\ref{eq:TTSH}) and using from incompressibilty condition in 2d we get, 
\begin{equation}
\frac{\partial \omega}{\partial t} + \lambda_0 \bm{u} \cdot \nabla \omega = -\alpha \omega - \beta \nabla \times \left( |\bm{u}|^2 \bm{u} \right)
+ \Gamma_0 \nabla^2 \omega - \Gamma_2 \nabla^4 \omega;
\end{equation}
where $\omega$ is $\nabla \times\textbf{u}$.\\
The simulations were performed on a square domain of size $160\times160$ with a uniform grid resolution of $4096\times4096$. This corresponds to a grid spacing of dx =  0.04, which sets the smallest resolvable length scale in the simulation.
This grid spacing is significantly smaller than the characteristic length scale, $\Lambda = 2\pi\sqrt{\frac{2\Gamma_2}{|\Gamma_0|}} = 0.4$. It is also smaller than the typical dissipation length scale, ensuring that small-scale structures and the fine details of the flow are well captured.
The choice of domain size was made to be much larger than the maximum velocity correlation length scale for the highest activity parameter $(\alpha = -9)$ used in our simulations. This ensures that the system can develop a broad range of scales without being artificially constrained by the domain boundaries and forming condensates. This setup allows the dynamics to evolve naturally, capturing both the large-scale collective motion and the finer-scale turbulent structures present in active systems. We have investigated the system for $\alpha \in [-9, -1]$. The root-mean-square velocity, $u_{\mathrm{rms}}$, is shown in Fig.~\ref{fig:vrms_spectra}(a). It increases linearly with $\alpha$ for $\alpha > \alpha_c$ \cite{cp2020friction}, but deviates from this linear trend for $\alpha < \alpha_c$, where $\alpha_c = -5$. The corresponding spectra are shown in Fig.~\ref{fig:vrms_spectra}(b), where the $-3/2$ slope emerges for $\alpha < \alpha_c$ \cite{mukherjee2023intermittency}.
We have used the FFTW module for Fourier transforms for the pseudo-spectral method.
The time marching is done using the second-order integrating factor method, where the stiff linear part is handled exactly through the integrating factor, improving stability. The simulations use a time step of $dt = 2\times10^{-4}$ for up to $2\times10^{5}$ iterations, corresponding to a total duration of $100\tau_{\Gamma}$, where $\tau_{\Gamma} =2\pi\sqrt{2}\frac{\Gamma_2}{|\Gamma_0|^2}$ is the instability growth timescale. Statistical averages are computed over the interval $[50\tau_{\Gamma},100\tau_{\Gamma}]$ using 50 snapshots.\\ 

Experiments on \textit{B.~subtilis}, at normal oxygen concentration, report mean swimming speeds of $\simeq 25~\mu\mathrm{m}/\mathrm{s}$ with a field of view of approximately $400~\mu\mathrm{m} \times 400~\mu\mathrm{m}$ ~\cite{wensink2012meso,cp2020friction,kiran2023irreversibility}. We map these experimental values onto the constant velocity scale $v_c$ and the simulation domain size $L \times L$, respectively. This mapping results in scale factors of $25/v_c$ for velocities and $4 \times 10^{-2}/L$ for lengths when relating our parameters to experimental observations. With this correspondence, the characteristic velocities $v_c$, $v$, and $u_{\mathrm{rms}}$ in the simulations correspond to $\simeq 25~\mu\mathrm{m}/\mathrm{s}$, $\simeq (25~\mu\mathrm{m}/\mathrm{s}$ to $75~\mu\mathrm{m}/\mathrm{s})$, and $\simeq (4.5~\mu\mathrm{m}/\mathrm{s}$ to $70~\mu\mathrm{m}/\mathrm{s})$, respectively. Likewise, the parameter $\Lambda$, which sets the characteristic linear size of vortical structures, corresponds to $\simeq 25~\mu\mathrm{m}$.

\begin{figure}[ht!]
    \centering
    \includegraphics[width=0.7\textwidth]{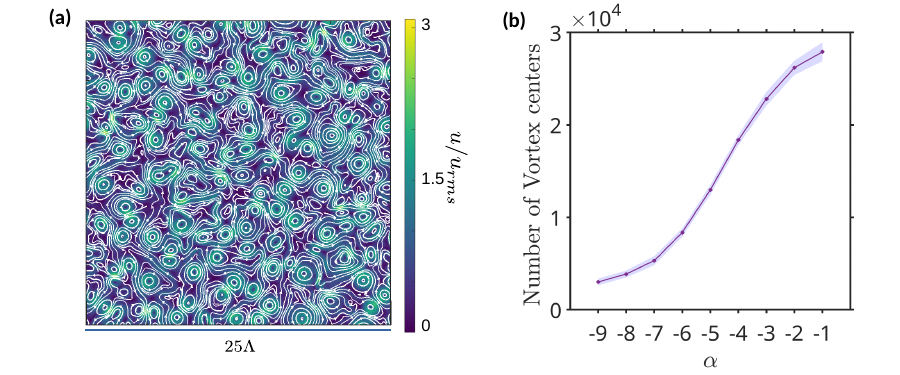}
    \caption{(a) Velocity streamlines overlaid on the velocity magnitude are shown for $\alpha = -1$ in a $25\Lambda \times 25\Lambda$ domain, highlighting flow structures predominantly governed by the length scale $\mathcal{O}(\Lambda)$. The color bar is normalized by $u_{\rm rms}$. (b) The time-averaged number of identified vortex centers as a function of $\alpha$, showing a decrease in the number of distinct vorticity centers with increasing activity. The shaded region represents the fluctuations (standard deviation) over time.}
    \label{fig:num_vor}
    \end{figure}

\section{Details of extraction of vortex centers}
\begin{figure}[h!]
  \includegraphics[width=19cm]{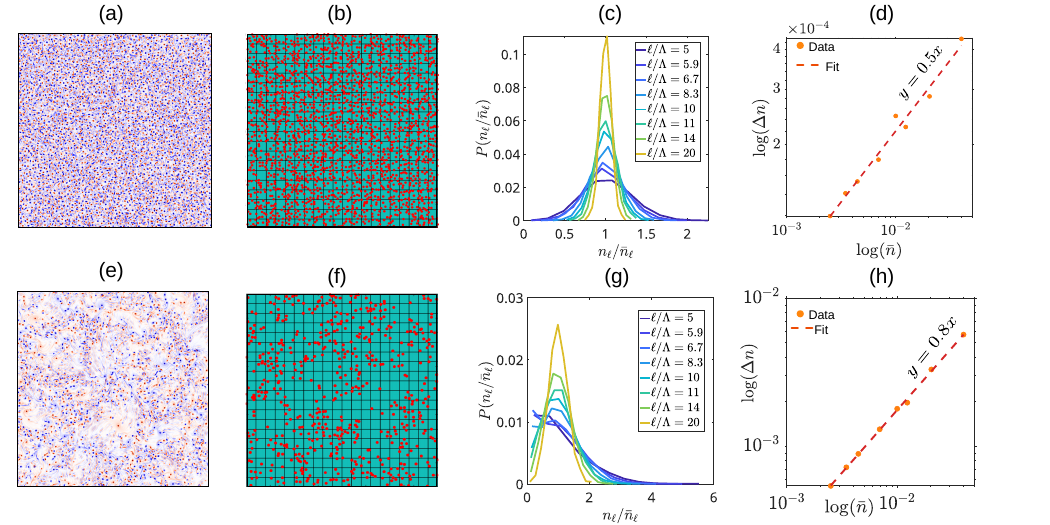}
  \caption{ (a) and (e) show the vorticity field at a representative zoomed-in snapshot for weak ($\alpha = -1$) and strong ($\alpha = -6$) activity, respectively, revealing a transition from spatially disordered vortex structures to highly structured patterns with dense vortical regions and zero or less vorticity regions. 
(b) and (f) display the corresponding extracted vortex centers, obtained by identifying regions below a threshold in the Okubo–Weiss parameter and computing their intensity-weighted centroids. 
(c) and (g) show the probability distribution functions of the number of vortex centers $n_\ell$ within subdomains of varying size $\ell$, with$\ell \in [5 \Lambda, 20 \Lambda$], highlighting the change in spatial homogeneity. 
(d) and (h) present the number fluctuations $\Delta n$ as a function of the average $\bar n_\ell$, respectively, for the two activity levels. For low activity ($\alpha = -1$), $\Delta n \sim \bar n_\ell ^{0.5}$, indicating homogeneous distribution, while for high activity ($\alpha = -6$), the scaling approaches $\Delta n \sim \bar n_\ell ^{0.8}$, consistent with inhomogenity in vortex centers distribution.
} 
  \label{vor_extract}
\end{figure}

\begin{figure}[h]
     \centering
     \includegraphics[width=0.4\textwidth]{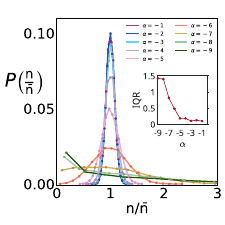}
     \caption{\small{Probability distribution functions (PDF) of the normalized vortex count $(\rm n /\bar{n})$ in subdomains of fixed linear size \(\ell\) for different activity values.
 As \(|\alpha|\) increases the distributions broaden and their maxima
 shift from \(\rm n /\bar{n} \simeq 1\) toward lower values, indicating that typical subdomains contain fewer vortex centers, while strong fluctuations become more likely. The inset shows the interquartile range (IQR) of the distribution. }}
     \label{fig:vor_distribution}
 \end{figure}

{\label{extract_vor_cenerts}}
For the number fluctuation analysis, from each snapshot, we extracted the intense vorticity regions by applying a threshold on the Okubo-Weiss parameter. The Okubo-Weiss parameter, defined as $Q = \frac{\omega^2-\sigma^2}{4}$, where $\sigma^{2} = \frac{1}{2}\sum_{ij}(\partial_i u_j+\partial_j u_i)^2$ with $i,j = 1,2$ can be used to determine the vorticity-dominated and strain dominated regions \cite{weiss1991dynamics,elhmaidi1993elementary}. We extracted the regions with $Q < Q_t$. Within each of these regions that are connected vorticity-dominated patches, we estimated the vortex center using an intensity-weighted centroid, where the vorticity field serves as the weighting function. If $( (x_i, y_i) )$ denote the coordinates of the $( i^{th} )$ grid in a region, and $( w_i)$ is the corresponding vorticity magnitude at that location, then the centroid $( (x_c, y_c) )$ is computed as:
\begin{equation}
x_c = \frac{\sum_i x_i \cdot w_i}{\sum_i w_i}, \quad
y_c = \frac{\sum_i y_i \cdot w_i}{\sum_i w_i}.
\end{equation}
The identified centroids are then used in further statistical analyses, such as computing number fluctuations across subdomains of varying size(discussed in the main text, see Fig.~\ref{vor_extract}), Voronoi analysis (discussed in the next section). 
 To analyze the distribution of intense vortices for each time step, the system is divided into subregions of area $\ell^2$, and estimate the number of vortex centers $n$ in each subregion, the mean number of vortex centers $\bar n$ and the standard deviation $\Delta n$, all normalized by the total number of vortex centers. The subregion size is varied over a range with $\ell \in [5\Lambda, 20\Lambda]$. In Fig.~\ref{fig:vor_distribution},
  we plot the probability distribution  function (PDF) of $\rm n$ around for different $\alpha$, in a given subregion. We note that for decreasing $\alpha$, the $P(\frac{\rm n}{\bar{n}})$ broadens, indicating stronger fluctuations in the number of vortex centers. The inset displays the interquartile range (IQR), which serves as a measure of the spread in the number distribution. The IQR is defined as the difference between the 75th percentile $(Q_3)$ and the 25th percentile $(Q_1)$ of the distribution: $$IQR = Q_3-Q_1$$ The IQR effectively captures the range within which the central $50\%$ of the data fall. 

\section{Static Structure factor}
\begin{figure}[h]
     \centering
     \includegraphics[width=0.7\textwidth]{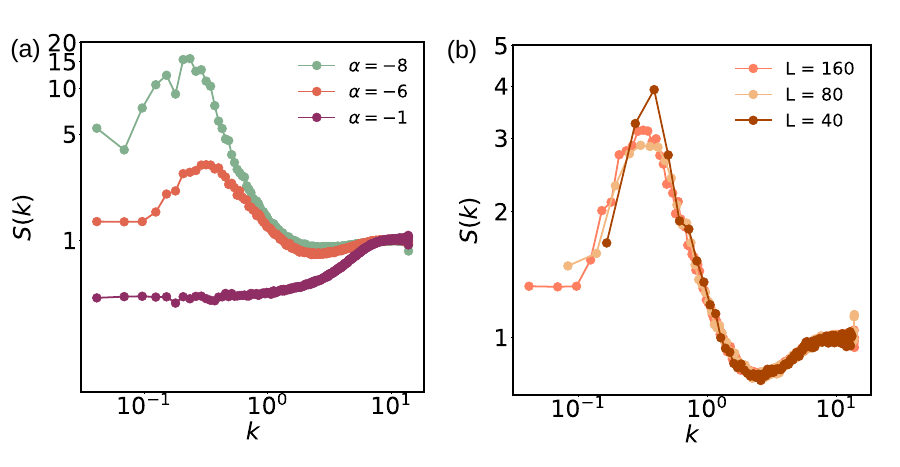}
     \caption{\small{(a) Static Structure factor for different activity. (b) Static Structure factor for fixed activity ($\alpha= -6$) and varying system size L.}}
     \label{fig:static_structure_factor}
 \end{figure}

{\label{structure factor}}
We compute the static structure factor of the number density of vortex centers, which characterizes equal-time spatial correlations in number density. To this end, the domain is coarse-grained into square boxes of size $l=5\Lambda$. The local density fluctuation is defined as \(\delta\rho_l = \rho_l - \rho_0\), where $\rho_l=n/l^2$, $n$ is the number of vortex centers in a given box of size $l^2$ and \(\rho_0\) is the mean density. The Fourier transform \(\delta\rho_l(\mathbf{k})\) is then used to compute the structure factor according to the standard definition~\cite{toner2005hydrodynamics,fily2012athermal}.
\begin{equation}
S(k) = \frac{1}{N}\left\langle |\delta\rho_l(\mathbf{k})|^2 \right\rangle .
\end{equation}
For large activities ($\alpha < \alpha_c$), the resulting structure factor \(S(k)\) increases with decreasing wave number \(k\) at intermediate scales, reaches a maximum at a finite \(k\), and subsequently decreases as \(k \to 0\). Consequently, \(S(k)\) remains finite in the long-wavelength limit, in contrast to the divergence observed in MIPS~\cite{fily2012athermal}. This behavior indicates that giant number fluctuations are present only over short to intermediate length scales (at mesoscale) and are suppressed at larger scales as shown for $\alpha = -6 $ and $-8$ in Fig~\ref{fig:static_structure_factor}(a). For smaller activities ($\alpha> \alpha_c$), where giant number fluctuations are absent, the structure factor does not have a characteristic peak and saturates to a constant value as $k\to0$, shown for $\alpha = -1$ in Fig~\ref{fig:static_structure_factor}(a). Furthermore, for a given value of $\alpha$ for example ($\alpha = -6$), the characteristic peak of the structure factor (seen at $\sim50\Lambda$) is unaffected by changes in system size $L$ (we vary system size between $100\Lambda-500\Lambda$, Fig~\ref{fig:static_structure_factor}(b)).

\section{Voronoi Analysis of Vortex Center Distribution}
\begin{figure}[h!]
  \includegraphics[width=13cm]{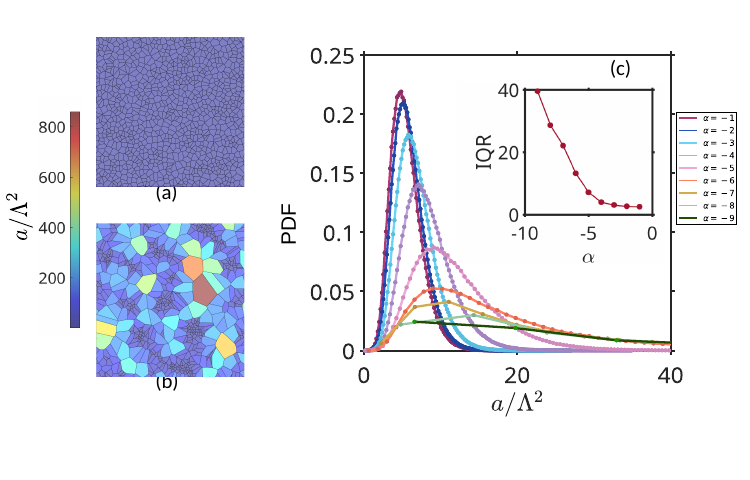}
  \caption{ (a) and (b) represent the area constructed by Voronoi tessellation based on the filtered vortex index for $\alpha = -1$ and $\alpha = -8$, respectively. (c) The probability distribution of the area normalised by $\Lambda^2$ of these Voronoi polygons. The inset shows the interquartile range (IQR) of the distribution.}
  \label{vornoi}
\end{figure}

{\label{voronoi_dist}}
To further quantify the spatial inhomogeneity in the distribution of vortex centers, we performed a Voronoi tessellation analysis. A Voronoi tessellation divides the domain into polygonal cells such that each cell contains exactly one vortex center, and every point in a given cell is closer to its associated center than to any other. This method provides a geometric way to measure the local spacing between vortex centers and is widely used to detect clustering or depletion in point patterns~\cite{osawa2021voronoi}.
By calculating the area of each Voronoi cell, we obtained a distribution of cell areas that serves as a proxy for the local density of vortex centers. In a spatially homogeneous system, Fig.~\ref{vornoi}(a), the distribution of Voronoi cell areas is expected to be narrow and peaked. However, for increasing activity, we observed a broadening of this distribution, Fig.~\ref{vornoi}(b), indicating the emergence of clustering and void-like regions, Fig.~\ref{vornoi}(c).
In particular, our results show that for activity levels below $(\alpha_c = -5)$, the Voronoi area distribution departs significantly from that Gaussian distribution. This transition reflects the development of spatial inhomogeneity in the vortex center positions, consistent with our findings from the number fluctuation analysis. The inset shows the IQR of area distribution increases with activity, indicating that the distribution of Voronoi areas becomes more spread out at higher activity levels.
\newpage
\section{Non-dimensionalization and polar reduction for an isolated vortex}{\label{Polar_form}}

\subsection{Dimensional model}
We begin with Toner–Tu–Swift–Hohenberg (TTSH) type equation for an incompressible active fluid,
\begin{equation}
\label{eq:dimensional}
\partial_t \mathbf u + \lambda_0\,(\mathbf u\!\cdot\!\nabla)\mathbf u
= -\nabla p \;+\; \Gamma_0 \nabla^2 \mathbf u \;-\; \Gamma_2 \nabla^4 \mathbf u
\;-\; \bigl(\alpha + \beta |\mathbf u|^2\bigr)\mathbf u
\end{equation}

\subsection{Choice of scales and non-dimensionalization}
Scaling length by $\Lambda = \sqrt{\frac{\Gamma_2}{|\Gamma_0|}}$, time with $\tau_{\alpha}=\frac{1}{|\alpha|}$, velocity with $v=\sqrt\frac{-\alpha}{\beta}$, and pressure accordingly:
\[
\mathbf x=\Lambda\,\mathbf x',\qquad
t=\frac{t'}{\alpha},\qquad
\mathbf u=v\,\mathbf u',\qquad
p=\alpha \Lambda v\,p'\qquad
\]
Under these, derivatives rescale as
\[
\nabla = \frac{1}{\Lambda}\nabla',\qquad
\partial_t = \alpha\,\partial_{t'}
\]
Substituting into \eqref{eq:dimensional} and dividing by $\alpha v$, gives
\begin{equation}
\label{eq:nd-primed}
\partial_{t'}\mathbf u' + \frac{\lambda_0 v}{\alpha \Lambda}\,(\mathbf u'\!\cdot\!\nabla')\mathbf u'
= -\nabla' p'
+ \underbrace{\frac{\Gamma_0}{\alpha \Lambda^2}}_{=\,R} \nabla'^2 \mathbf u'
- \underbrace{\frac{\Gamma_2}{\alpha \Lambda^4}}_{=\,R} \nabla'^4 \mathbf u'
- \Bigl(1-|\mathbf u'|^2\Bigr)\mathbf u'.
\end{equation}
where, 
\[
 R \; = \;\frac{\Gamma_0}{\alpha \Lambda^2}=\frac{\Gamma_2}{\alpha \Lambda^4}
= \frac{\Gamma_0^2}{\alpha \Gamma_2} 
\]
and dropping primes, the non-dimensional system is
\begin{equation}
\label{eq:nd-final}
\partial_t \mathbf u + \frac{\lambda_0 v}{\alpha \Lambda}\,(\mathbf u\!\cdot\!\nabla)\mathbf u
= -\nabla p \;+\; R\,\bigl(\nabla^2-\nabla^4\bigr)\mathbf u
\;-\; \bigl(1-|\mathbf u|^2\bigr)\mathbf u .
\end{equation}

\subsection{ Steady azimuthal vortex ansatz in polar coordinates}
We consider a steady, purely azimuthal, axisymmetric vortex in 2D polar coordinates $(r,\theta)$:
\[
\partial_t \mathbf u=0,\qquad
\mathbf u(r)=u_r\,\hat{\mathbf r}+u_\theta\,\hat{\boldsymbol\theta},\qquad
u_r(r)=0,\quad u_\theta(r)=f(r).
\]
With $p=p(r)$ by symmetry, the $\theta$–component of the advective term and of $\nabla p$ vanish.
The vector Laplacian acting on the $\theta$–component (no $\theta$-dependence) reads
\begin{equation}
\label{eq:L-def}
(\nabla^2 \mathbf u)_\theta \;=\; \mathcal L[f]
:= f'' + \frac{1}{r} f' - \frac{f}{r^2}.
\end{equation}
Applying the same (componentwise) Laplacian again gives the bilaplacian contribution
\begin{equation}
\label{eq:L2-def}
(\nabla^4 \mathbf u)_\theta \;=\; \mathcal L\!\bigl[\mathcal L[f]\bigr]
= f^{(4)} + \frac{2}{r} f^{(3)} - \frac{3}{r^2} f'' + \frac{3}{r^3} f' - \frac{3}{r^4} f.
\end{equation}
(Equations \eqref{eq:L-def}–\eqref{eq:L2-def} follow by straightforward differentiation of $\mathcal L$.)

\subsection{ Reduced fourth-order ODE for the vortex profile}
Taking the $\theta$–component of \eqref{eq:nd-final} under the above ansatz yields
\begin{equation}
\label{eq:ODE-compact}
R\,\bigl(\mathcal L^2 f -\mathcal L f \bigr) \;=\; -(1-f^2)\,f .
\end{equation}
Expanding with \eqref{eq:L-def}–\eqref{eq:L2-def} gives the explicit radial ODE
\begin{equation}
\label{eq:ODE-expanded}
R\left[
f^{(4)} + \frac{2}{r} f^{(3)}
- \left(1 + \frac{3}{r^2}\right) f''
+ \left(\frac{3}{r^3}-\frac{1}{r}\right) f'
+ \left(\frac{1}{r^2} - \frac{3}{r^4}\right) f
\right] \;=\; -(1 - f^2)\,f .
\end{equation}

\subsection{Small-$r$ asymptotics}

For $r \ll \Lambda$, we assume a power-law ansatz for the azimuthal velocity,
\begin{equation}
f(r) \sim r^a .
\end{equation}
Substituting this form into Eq.~\ref{eq:ODE-expanded} and retaining the dominant 
$\mathcal{O}\!\left(\tfrac{1}{r^4}\right)$ contributions yields an algebraic equation for $a$. 
\begin{equation}
(a-1)^2(a-3)(a+1) =0 
\end{equation}
The distinct roots are
\begin{equation}
a = 1, \qquad a = 3, \qquad a = -1 
\end{equation}
The negative exponents is discarded since it leads to divergences in $f(r)$. 
The cubic branch is regular but subdominant, so the small-$r$ behaviour is dominated by the linear solution
\[
f(r) \sim A r .
\]
\begin{figure} [h!]
    \centering
    \includegraphics[width=0.4\textwidth]{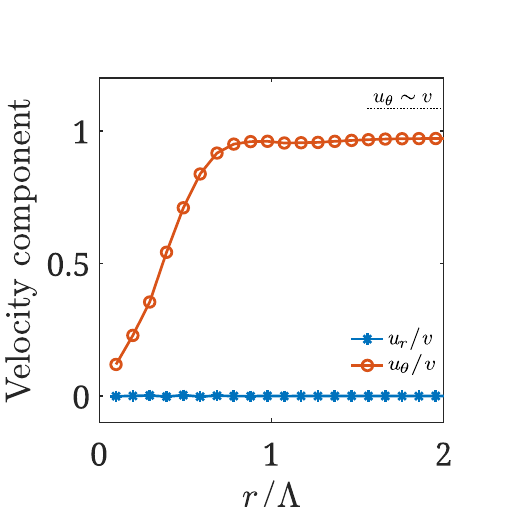}
    \caption{\small{Radial ($u_r$) and azimuthal ($u_{\theta}$) velocity components near an isolated vortex point as a function of radius $r$. The profiles show that the velocity is primarily aligned in the azimuthal direction with negligible radial component. For $r > \Lambda$ (asymptotic limit), the azimuthal velocity converges to $v = \sqrt{|\alpha|/\beta}$. }}
    \label{fig:vel_comp_alpha_7}
    \end{figure}

\section{Order Parameter}{\label{order_param}}
We explored additional parameter regimes by varying the strength of the higher-order term $\Gamma_2$, which modulates the influence of the Swift-Hohenberg instability. Our results underscore the critical role of the interplay between two competing physical mechanisms: the Toner-Tu term, which promotes local alignment and polar order, and the Swift-Hohenberg term, which favors spatial modulation and destabilizes uniform configurations.

\begin{figure}[h!]
  \includegraphics[width=1.0\textwidth]{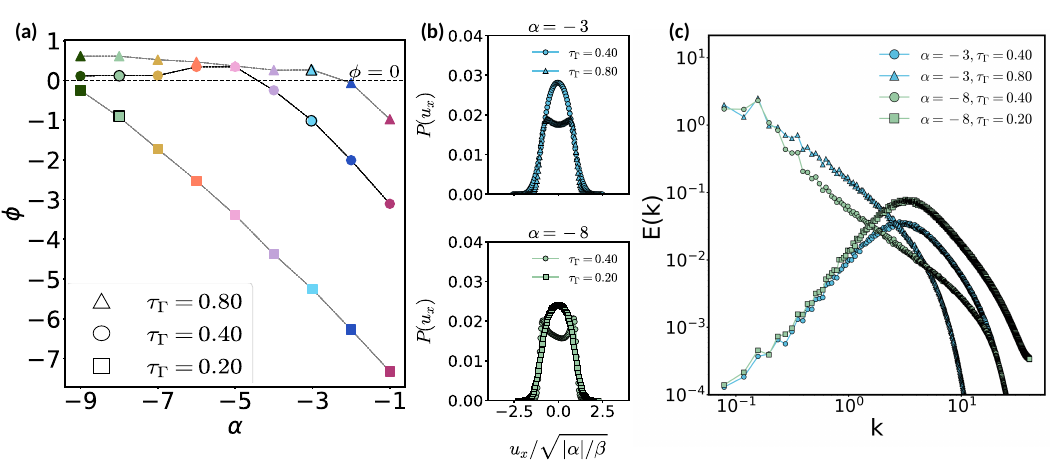}
  \caption {((a) Order parameter $\phi$ as a function of activity $\alpha$ for three different values of $\tau_{\Gamma}$ obtained by varying $\Gamma_2$. The curves highlight how the critical activity at which the transition in flow structure occurs shifts with $\tau_{\Gamma}$. 
(b, top) Probability distribution function of normalized velocity component,  for $\alpha = -3$, demonstrating a transition from a unimodal Gaussian profile to a bimodal form as $\phi$ becomes positive, marking the onset of local ordering. The points highlighted with black borders indicate the cases for which the velocity component distributions are shown in panel (b). (b, bottom) Corresponding normalized velocity component distribution for $\alpha = -8$, illustrating the reverse transition from a bimodal to a Gaussian profile as local order breaks down.
(d) Energy spectra for the two representative cases: at $\alpha = -8$, the spectrum changes from a $k^{-3/2}$ power law to no clear scaling as the system enters a disordered state; conversely, at $\alpha = -3$, a $k^{-3/2}$ scaling emerges from an initially structureless spectrum, reflecting the onset of ordering.
}
  \label{order_param_add}
\end{figure}

This competition governs the emergence and breakdown of mesoscale structures. For example, at a fixed activity level of $\alpha = -3$, increasing $\Gamma_{\tau}$ from 0.40 to 0.80 by increasing  $\Gamma_2$, induces a transition from a disordered to a locally ordered state, as indicated by the change in the sign of the order parameter ($\phi < 0 \to \phi > 0$) and a corresponding shift in the velocity distribution from unimodal to bimodal Fig.~\ref{order_param_add}(b). Conversely, at $\alpha = -8$, decreasing $\Gamma_{\tau}$ to 0.20 from 0.40 by decreasing $\Gamma_2$, destabilizes the ordered state, leading to $\phi$ switching from positive to negative and the velocity statistics reverting to a unimodal form Fig.~\ref{order_param_add}(b). Correspondingly, the energy spectra exhibit distinct changes, reflecting the redistribution of energy across scales as the system transitions between ordered and disordered regimes, Fig.~\ref{order_param_add}(c).

These findings demonstrate that the onset of local ordering is not governed solely by activity strength, but emerges from a balance between active alignment and the intrinsic instability of the system. The velocity distribution and order parameter $\phi$ serve as sensitive diagnostics of this balance, capturing the transition in the system’s collective dynamics.

\bibliographystyle{unsrt}

\end{document}